\documentclass[12pt]{article}
\usepackage{amsmath, amssymb}
\begin{document}

\begin{center}
{\bf \large{Nonclassical Total Probability Formula and Quantum Interference of
Probabilities}}
\end{center}
\vskip0,5cm
\begin{center}
{\bf Alexander Bulinski}$^{a}$,\;\;{\bf Andrei Khrennikov}$^{b}$

\end{center}
\vskip0,2cm

$^{a}${\it Department of Mathematics and Mechanics,

Moscow State University, Moscow 119992, Russia}

$^{b}${\it International Centre for Mathematical Modeling in Physics and
Cognitive Sciences,

MSI, University of V\"axj\"o, S-35195 Sweden}
\vskip0,3cm

{\it E-mail addresses:}
\vskip0,1cm
bulinski@mech.math.msu.su (A.V.Bulinski),

Andrei.Khrennikov@msi.vxu.se (A.Yu.Khrennikov)

\vskip0,4cm
{\bf Abstract}

Interpretation of the nonclassical total probability formula
arising in some quantum experiments is provided based
on stochastic models described by means of a sequence of random
vectors changing in the measurement procedures.

\vskip0,3cm
{\it MSC:} Primary: 60F15
\vskip0,3cm
{\it Keywords:} the total probability formula, statistical ensembles of
systems, frequences, pairwise independent random vectors, SLLN
\vskip1cm

{\bf 1. Introduction}

\vskip0,5cm
We are interested in giving meaning, within Kolmogorov's axiomatics, to
nonclassical versions of the total probability formula.
This problem was considered in Khrennikov, 2001a, 2001b,
on physical level of rigor, using the frequency approach, see von Mises,
1964.

Almost any treatise on quatum mechanics dwells on the fundamental two-slot
experiment with e.g. electrons passing through one or two open slits (see,
e.g.,
d'Espagnat, 1999, p.6). Let $A_i$ denote an event of passing through the
slit with label $i$, here $i = 1,2$. Interpretation of the results of this
experiment and the related ones (see, e.g., Feynman and Hibbs, 1965, p.11)
has led to the following
formula for the probability of electron passing through the two open slots:
\begin{equation}\label{1}
P(A_1 \cup A_2) = P(A_1) + P(A_2) +
2 \sqrt{P(A_1) P(A_2)} \cos \theta
\end{equation}
where $P$ is a symbol of probability and $\theta$ a certain parameter.
Usually the origin of the term
$2 \sqrt{P(A_1) P(A_2)} \cos \theta$
is explained by reffering to "self-interference" inherent to the "wave nature"
of an electron.
To be more precise, calculating relative frequencies of the considered events
after repetitions of the experiment led to conclusion that the classical
formula of addition for probabilities of disjoint events should be modified.

A remarkble achievement of physics was the employment of Hilbert space methods
for description of quantum mechanical systems. In the standard formalism (see,
e.g., d'Espagnat, 1995, p. 63-65) one derives formula \eqref{1} with the
help of superposition
principle.  This elegant proof using complex-valued coefficients of a
"state-vector" decomposition with respect to the basis of eigenvectors of a
self-ajoint operator corresponding to certain "observable" rather loses links
with traditional concepts of probability theory. Moreover, a number of
specialists in quantum mechanics believe that with the Kol\-mo\-go\-rov
axiomatics it is impossible to justify formula \eqref{1}. The discussion
continues from the early days of the subject and still the joint view-point is
not generated, see, e.g., Heisenberg, 1930, Dirac, 1930, Bohr, 1934,
von Neumann, 1955,
Feynman and Hibbs, 1965, Jauch, 1968, Beltrametti and Cassinelli, 1981,
Holevo, 1980, 2001, Accardi, 1984, Ballentine, 1986, Parthasaraty, 1992,
Meyer, 1993, Peres, 1994, d'Espagnat, 1995, 1999, Busch et al., 1995.

The careful analysis of probability spaces and families of random variables
used for models  of stochastic experiments gives the key to understanding
the origin of the formulas like \eqref{1} (ch. Khrennikov, 1999, p.100-102,
2000, p. 5936,Thorisson, 2000, p.29, Accardi and Regoli, 2001, p.8-9).
The main role here is played by the following {\it contextualism
principle}\footnote{In quantum theory
this principle was introduced by N. Bohr who underlined that experimental
conditions
plays the crucial role in determining physical observables, see e.g. Bohr,
1934.}:
introduction of a random variable or a collection of random variables
describing some properties of a physical system or an ensemble of systems
should take into account the total
\footnote{The choice of the principal features (conditions) leads to various
models.}
collection (class) of the conditions, i.e. {\it context},
under which the values
of these variables (or their distributions) were specified.

In fact the contextualism principle can also be found in the fundamental work
by Kol\-mo\-go\-rov, 1933, or in his paper, 1965 (p. 249):
"Thus, to say that an event $A$ is "random" or "stochastic" and to assign
it a definite probability $p = P(A|{\cal S})$ is possible only when we have
already determined the class of allowable ways of setting up the series of
experiments.  The nature of this class will be assumed to be included in the
conditions ${\cal S}$". So, the role of a specified class of conditions which
can be reproduced is stressed (see details in Gnedenko, 1962, Shiryaev, 1998).

The aim of this paper is to show that a simple stochastic model of
measurement of characteristics of some physical systems provides a quite
natural explanation for the "interferentional" term arising in the formula
\eqref{1}.  Let us remark once more that besides the probability space
$(\Omega,{\cal F},P)$ the essential role in construction of probabilistic
models is played by families of random variables used for description of
studied phenomena. Of course, having specified the model, one can
consider the new ({\it canonical}) probability space with coordinate random
variables.

It should be emphasized that neither
we are going to analyze concrete quantum
mechanical systems, nor impose on them a classical probability space
$(\Omega,{\cal F},P)$. Our goal is not to transform the physical phenomena
into "classical" probability events, i.e. elements of a $\sigma$-algebra
$\cal F$ of subsets of a space $\Omega$, where $P$ is a measure defined on
events and normalized so that $P(\Omega) =1$.
Deep study of the quantum probability problems can be found, e.g., in
Accardi, 1984,2001, Holevo, 1980, 2001, Meyer, 1993, Parthasaraty, 1992.

In the sequel it will be convenient to give another form to relation
\eqref{1}.
Set $C = A_1 \cup A_2$ where $A_1 \cap A_2 = \varnothing$ and rewrite
\eqref{1}
as a "nonclassical total probability formula":
\begin{equation}\label{2}
P(C) = P(C|A_1)P(A_1) + P(C|A_2)P(A_2) + 2\sqrt{P(C|A_1)P(A_1)P(C|A_2)P(A_2)}
\cos \theta,
\end{equation}
where, as usual,
$$
P(C|A_i) = P(CA_i)/P(A_i)
$$
and $P(A_i) > 0$, $i=1,2$. The trivial case $P(A_i) = 0$ for some $i$ is
henceforth excluded.

\vskip1,5cm
{\bf 2. The model description}
\vskip0,5cm
For the sake of simplicity consider some physical system $S$ posessing only
properties \footnote{On the notion of a property see, e.g. Khrennikov, 1999,
p.  59.} {\bf A} and {\bf C} described by a pair of two-valued random
variables ({\bf A} and {\bf C} "characteristics"). Suppose that there is a
procedure permitting to reproduce the copies $S_1,S_2,\ldots$ of $S$ to get a
{\it statistical ensemble} of $S$ in the following sense. Let
$(a^{(1)},c^{(1)}),(a^{(2)},c^{(2)}),\ldots$ be a sequence of i.i.d. random
vectors defined on some probability space $(\Omega,{\cal F},P)$. Assume
that for some $a_i,c_i \in \mathbb{R}, i=1,2$
\begin{equation}\label{0}
P(a^{(1)} = a_1) = p^a_1,\;\;\;P(a^{(1)} = a_2) = p^a_2, \end{equation}
\begin{equation}\label{0'} P(c^{(1)} = c_1) = p^c_1,\;\;\;P(c^{(1)} = c_2) =
p^c_2, \end{equation} where $p^a_1 + p^a_2 =1$ and $p^c_1 + p^c_2 =1$.

For $j \in \mathbb{N}$ and elementary event $\omega \in \Omega$ the vector
$(a^{(j)}(\omega),c^{(j)}(\omega))$ characterizes the properties {\bf A}
and {\bf C} of the system $S_j$. The independence assumption for the sequence
$\{(a^{(j)},c^{(j)})\}_{j \geq 1}$ reflects the usual noninteraction
hypothesis
for elements of stochastic ensemble. Further we can use more general
assumptions as well. For each $j \in \mathbb{N}$ no conditions are imposed on
the joint distribution of random variables $a^{(j)}$ and
$c^{(j)}$.\footnote{This is very natural
from quantum viewpoint.} Moreover,
note that for any $\omega \in \Omega$ the sequence
$\{(a^{(j)}(\omega),c^{(j)}(\omega)\}_{j \geq 1}$ is not, in general,
available. In fact, to get the data certain measurement procedures should
be used.

Suppose that it is possible to apply to systems $S_j$ a measurement procedure
${\cal M}_{\bf A}$
permitting for each $\omega \in \Omega$ to fix the value  $a^{(j)}(\omega)$
of the characteristic describing property {\bf A}. In other words, for every
$\omega \in \Omega$ we have integers $1 \leq k_1(\omega) < k_2(\omega) <
\ldots$ such that
$$
a^{(k_1(\omega))}(\omega) = a_1,\;\;
a^{(k_2(\omega))}(\omega) = a_1,\ldots
$$
and, respectively,
$$
a^{(m_j(\omega))}
(\omega) = a_2\;\;\mbox{for}\;\;m_1(\omega) < m_2(\omega)
<\ldots
$$
where $m_j(\omega) \in \mathbb{N} \setminus \cup_i \{k_i(\omega)\}$.

It is worth to note that after the procedure ${\cal M}_{\bf A}$ the property
{\bf C} of initial systems $S_j$ in general can be changed. Thus, new systems
$S'_j$ $(j \in \mathbb{N})$ arise. To describe the property {\bf C} of these
systems it is natural to use random sequences
$\{\overline{c}^{(j)}\}_{j \geq 1}$
and
$\{\widehat{c}^{(j)}\}_{j \geq 1}$
such that
\begin{equation}\label{0a}
P(\overline{c}^{(j)} = c_r) = p^{\overline{c}}_r,\;\;
P(\widehat{c}^{(j)} = c_r) = p^{\widehat{c}}_r,\;\;\;j,r = 1,2,
\end{equation}
where
$p^{\overline{c}}_1 + p^{\overline{c}}_2 = 1$,
$p^{\widehat{c}}_1 + p^{\widehat{c}}_2 = 1$.

Assume that for each $\omega \in \Omega$ and $i = i(\omega) \in \mathbb{N}$
such that $a^{(i)}(\omega) = a_1$
the value $\overline{c}^{(i)}(\omega)$ appears instead of $c^{(i)}(\omega)$.
Other values $c^{(i)}(\omega)$ (for $i = i(\omega) \in \mathbb{N}\setminus
\cup_i \{k_i(\omega)\}$) are replaced by $\widehat{c}^{(i)}(\omega)$.

So, to describe the properties {\bf A} and {\bf C} for the ensemble of systems
$S'_j$ $(j \in \mathbb{N}$) we obtain two-dimensional sequences
$$
\{(a_1,\overline{c}^{(k_j(\omega))} (\omega))\}_{j \geq1}\;\;\;\mbox{and}
\;\;\; \{(a_2,\widehat{c}^{(m_j(\omega))} (\omega))\}_{j \geq1}
$$
Note that since $k_j = k_j(\omega)$ and $m_j = m_j(\omega)$ are random
variables defined on probability space $(\Omega,{\cal F},P)$ the same is
true of
$\overline{c}^{(k_j)} = \overline{c}^{(k_j(\omega))}(\omega)$ and
$\widehat{c}^{(m_j)} = \widehat{c}^{(m_j(\omega))}(\omega)$.

It is always possible to construct a probability space $(\Omega,{\cal F},P)$
where stochastic sequences $\{(a^{(j)},c^{(j)}\}_{j \geq 1}$,
$\{\overline{c}^{(j)}\}_{j \geq 1}$, $\{\widehat{c}^{(j)}\}_{j \geq 1}$
with the above-mentioned properties are defined. Moreover, using the standard
product of probability spaces it is easy to guarantee the independence of
terms of these sequences and independence of the collection of three sequences
as well. However, we do not use the last opportunity.

Suppose also that a measurement procedure ${\cal M}_{\bf C}$ can be applied
to systems $S'_j$ $(j \in \mathbb{N}$) to fix the values of the
characteristics
of property {\bf C}.
Thus we get systems $S_i^{"}$ $(i \in \mathbb{N}$) having
$\overline{c}^{(k_j(\omega))}(\omega)$ or $\widehat{c}^{(m_j(\omega))}
(\omega)$ as {\bf C}-characteristics for $\omega \in \Omega$ and $i =
k_j(\omega)$ or $i = m_j(\omega)$. Of course, the {\bf A}-characteristics
of systems $S'_j$ $(j \in \mathbb{N}$) in general can not be conserved.

\vskip1,5cm
{\bf 3. Nonclassical total probability formula}
\vskip0,5cm
Now we consider from the view-point of limit theorems for random variables
defined on $(\Omega,{\cal F},P)$ the approach used in Khrennikov,  2001a, 2001b.
Namely, we discuss statistical properties of {\bf C}-characteristics for
ensembles of systems $\{S_j\}_{j \geq 1}$ and $\{S_j^{"}\}_{j \geq 1}$. In
this
regard it is useful to recall the following statement from d'Espagnat, 1999,
(see p. 15): "Quantum mechanics is essentially a statistical theory. Except
in special cases it makes no prediction that bears on individual systems.
Rather, it predicts statistical frequencies. In other words, it predicts
as a rule, the number $n$ of times that a given event will be observed when
a large number $N$ of physical systems of the same type and satisfying
specified conditions ar esubjected to a given measurement process".

We are not going to analyze the measurement procedures, prepearing (for
measurement) procedures and the mechanism govering the modification of initial
properties of systems under various influences (see, e.g., Holevo, 1980,
2001).

Our very simple model with changing random variables permits to explain easily
the asymptotical behaviour for frequences of events studied in Khrennikov,
 2001a, 2001b.

Introduce some notation. Let $|\cdot|$ be a number of elements of a finite
set. For $\omega \in \Omega$, $N \in \mathbb{N}$ and $j = 1,2$ let
\begin{equation}\label{2'}
q_{jN} (\omega) = |\{i \in \{1,\ldots,N\}: \;c^{(i)}(\omega) = c_j\}|/N.
\end{equation}
Set $N_1(\omega) = \max\{i:\;k_i(\omega) \leq N\}$, $N_2(\omega) = \max
\{i:\;m_i(\omega) \leq N\}$. For $\omega \in \Omega$ and $j=1,2$ put
$$
n_{j1}(\omega) = |\{i \in \{1,\ldots,N_1(\omega)\}:
\;c^{(k_i(\omega))}(\omega)
= c_j\}|,
$$
$$
n_{j2}(\omega) = |\{i \in \{1,\ldots,N_2(\omega)\}:
\;c^{(m_i(\omega))}(\omega)
= c_j\}|,
$$
$$
m_{j1}(\omega) = |\{i \in \{1,\ldots,N_1(\omega)\}: \;
\overline{c}^{(k_i(\omega))}(\omega)
= c_j\}|,
$$
$$
m_{j2}(\omega) = |\{i \in \{1,\ldots,N_2(\omega)\}: \;
\widehat{c}^{(m_i(\omega))}(\omega)
= c_j\}|.
$$
To simplify the notation (for $j,r=1,2$ and $\omega \in \Omega$) we omit the
dependence of $n_{jr}(\omega)$ and $m_{jr}(\omega)$ on $N$.

Evidently, for all $\omega \in \Omega$, $j =1,2$ and $N \in \mathbb{N}$ one
has
\begin{equation}\label{3}
q_{jN}(\omega) =
\frac{m_{j1}(\omega)}{N} +
\frac{m_{j2}(\omega)}{N} + \gamma_{jN}(\omega)
\end{equation}
where
$$
\gamma_{jN}(\omega) =
\frac{n_{j1}(\omega) - m_{j1}(\omega)}{N} +
\frac{n_{j2}(\omega) - m_{j2}(\omega)}{N}.
$$

\vskip0,2cm
{\bf Theorem 1}. Let $\{(a^{(j)},c^{(j)})\}_{j \geq
1},\;\{(a^{(j)},\overline{c}^{(j)})\}_{j \geq 1}$
and $\{(a^{(j)},\widehat{c}^{(j)})\}_{j \geq 1}$ be sequences of pair wise
independent random vectors identically distributed within each sequence,
i.e. for all $j \in \mathbb{N}$
$$
(a^{(j)},c^{(j)})
\stackrel{{\cal D}}{=} (a^{(1)},c^{(1)}),\;
(a^{(j)},\overline{c}^{(j)})
\stackrel{{\cal D}}{=} (a^{(1)},\overline{c}^{(1)}),\;
(a^{(j)},\widehat{c}^{(j)})
\stackrel{{\cal D}}{=} (a^{(1)},\widehat{c}^{(1)}),\;
$$
here ${\cal D}$ means the equality of distributions. Assume that
\eqref{0'} -- \eqref{0a} hold true and $a^{(1)}$ is not degenerate.
Then for all $j,r = 1,2$ and $P$-a.e $\omega \in \Omega$ there exist
nonrandom limits
\begin{equation}\label{4}
p_{jr}^{c|a}:=\lim\limits_{N \to \infty}
\frac{m_{jr}(\omega)}{N_r(\omega)},\;\;\;
\gamma_j:=\lim\limits_{N \to \infty} \gamma_{jN}(\omega)
\end{equation}
and besides
\begin{equation}\label{5}
p_j^c = p_{j1}^{c|a} p_1^a + p_{j2}^{c|a} p_2^a + \gamma_j.
\end{equation}
{\it Proof.} Borelean functions of pairwise indepndent random vectors are
pairwise independent. Therefore, by the Etemadi SLLN (see e.g. Borkar,
1995, p. 69)
one has for $P$-a.e. $\omega \in \Omega$ and $j =1,2$
\begin{equation}\label{6}
q_{jN}(\omega) = \frac{1}{N} \sum\limits_{i=1}^{n}
I_{\{c^{(i)}(\omega)=c_j\}} (\omega) \to E \,
I_{\{c^{(1)}=c_j\}}=p_j^c\;\;\mbox{as}\;\;N \to \infty,
\end{equation}
where the indicator function of an event $D$
$$
I_D (\omega) = \begin{cases}
1,&\omega\in D;\\
0,&\omega\notin D.
\end{cases}
$$
Analogously with probability one
\begin{equation}\label{7}
\frac{N_1(\omega)}{N} \to p_1^a\;\;\mbox{and}\;\;\frac{N_2(\omega)}{N} \to
p_2^a\;\;\;\mbox{as}\;\;\;N \to \infty.
\end{equation}
It is clear that
$$
\frac{m_{j1}(\omega)}{N} = \frac{1}{N}\sum\limits_{i=1}^{N}
I_{\{\overline{c}^{(i)}(\omega) = c_j\}}(\omega) I_{\{a^{(i)}(\omega) =
a_1\}} (\omega),
$$
$$
\frac{m_{j2}(\omega)}{N} = \frac{1}{N}\sum\limits_{i=1}^{N}
I_{\{\widehat{c}^{(i)}(\omega) = c_j\}}(\omega) I_{\{a^{(i)}(\omega) =
a_2\}} (\omega).
$$
The same reasoning shows that for $P$-a.e. $\omega \in \Omega$ there exist
\begin{equation}\label{8}
\lim\limits_{N \to \infty} \frac{m_{j1}(\omega)}{N} = P(\overline{c}^{(1)}
=a_j,a^{(1)} = a_1),
\end{equation}
\begin{equation}\label{9}
\lim\limits_{N \to \infty} \frac{m_{j2}(\omega)}{N} = P(\widehat{c}^{(1)}
=a_j,a^{(1)} = a_1).
\end{equation}
The event consisting of those $\omega$ rendering
$a^{(i)}(\omega) = a_1$ for all $i \in \mathbb{N}$ or $a^{(i)}(\omega) =
a_2$ for all $i \in \mathbb{N}$ has $P$- measure zero. So, there exists
an event $U$ with $P(U) = 1$ such that both $N_1(\omega) >0$ and
$N_2(\omega) > 0$ for all $\omega \in \Omega$ and $N > M(\omega)$. Hence
for $\omega \in U$ and $N > M(\omega)$ we have
$$
\frac{m_{jr}(\omega)}{N} = \frac{m_{jr}(\omega)}{N_r(\omega)}
\frac{N_r(\omega)}{N},\;\;\;j,r=1,2.
$$
Taking into account \eqref{7} -- \eqref{9} we establish the existence of
limits $p_{jr}^{c|a}$ in \eqref{4} and due to \eqref{6} the
validity of the second relation in \eqref{4} follows. Now \eqref{5} is obvious
and the proof is complete.

{\bf Remark 1.} Successive employment of procedures ${\cal M}_{\bf A}$ and
${\cal M}_{\bf C}$ for systems $S_j$ and $S'_j$ $(j \in \mathbb{N})$ permits
to determine $N_1(\omega)$, $N_2(\omega)$ and $m_{jr}(\omega)$ for $j,r=1,2$.
Thus, according to \eqref{4} and \eqref{7} there are strongly consistent
estimates for $p_{jr}^{c|a}$ and $p_r^a$ $(j,r=1,2)$. To determine
$q_{jN}(\omega)$ one should apply the procedure ${\cal M}_{\bf C}$ to systems
$S_1,\ldots,S_N$.
However, it would change the (distribution of) {\bf A}-characteristics of
these systems. Therefore, to obtain the strongly consistent estimates for
parameters $p_j^c$ $(j =1,2$) we can proceed in the following manner. Invoking
the principle of reproducibility we can construct besides the copies $S_j$
$(j \in \mathbb{N})$ of a system $S$ the copies $\widetilde{S}_j$ $(j
\in \mathbb{N})$ described by a sequence of random vectors
$\{(\widetilde{a}^{(j)},\widetilde{c}^{(j)})\}_{j \geq 1}$. The standard
extension of a probability space $(\Omega,{\cal F},P)$ can be used to define
this auxiliary random sequence. For an extension we keep the same notation
$(\Omega,{\cal F},P)$. Applying the procedure ${\cal M}_{\bf C}$ to systems
$\widetilde{S}_j$ $(j \in \mathbb{N})$ and defining
$\widetilde{q}_{jN}(\omega)$ analogously to \eqref{2'} we see that for
$P$-a.e.  $\omega \in \Omega$ and $j=1,2$ $$
\widetilde{q}_{jN}(\omega)/N \to p_j^c\;\;\;\mbox{as}\;N \to \infty.
$$ Thus under the conditions of Theorem 1 one can provide the strongly
consistent estimates for the values $\gamma_j$ $(j=1,2)$ appearing in formula
\eqref{5}.

{\bf Remark 2.} The notation $p_{jr}^{c|a}$ for limits in \eqref{4} seems
natural as $m_{jr}(\omega)/N_r(\omega)$ is the relative frequency of appearing
value $c_j$ for the characteristics of property {\bf C} in a sequence of
those systems which have $a_r$ value for the characteristics of property
{\bf A}. However, the sense of formula \eqref{5} is demonstrated much better
by relations \eqref{8} and \eqref{9} showing that

\begin{equation}\label{10}
p_{jq}^{c|a} = P(\overline{c}^{(1)} = c_j|a^{(1)} = a_1),\;\;p_{j2}^{c|a}
= P(\widehat{c}^{(1)} = c_j|a^{(1)} = a_2).
\end{equation}
Thus the additional (with respect to the classical total probability
formula) term $\gamma_j$ appears in \eqref{5} because the following equalities
need not hold true
$$
P(\overline{c}^{(1)} = c_j|a^{(1)} = a_1) =
P(c^{(1)} = c_j|a^{(1)} = a_1),
$$
$$
P(\widehat{c}^{(1)} = c_j|a^{(1)} = a_2) =
P(c^{(1)} = c_j|a^{(1)} = a_2),
$$
In such a way it is more informative to rewrite \eqref{5} in the form
\begin{equation}\label{11}
p_j^c = p_{j1}^{\overline{c}|a} p_1^a + p_{j2}^{\widehat{c}|a} p_2^a +
\gamma_j,
\end{equation}
setting now
\begin{equation}\label{12}
p_{j1}^{\overline{c}|a} = P(\overline{c}^{(1)} = c_j|a^{(1)} = a_1),\;
p_{j2}^{\widehat{c}|a} = P(\widehat{c}^{(1)} = c_j|a^{(1)} = a_2),\;\;j=1,2.
\end{equation}
Formula \eqref{11} clarifies as well that simultaneous measurement of
characteristics {\bf A} and {\bf C} in general is not supposed.

{\bf Remark 3.} Due to \eqref{11} and \eqref{12} one has
$$
|\gamma_j| \leq 1,\;\;j=1,2.
$$
In the particular case when for each $j \in \mathbb{N}$ random variables
$\overline{c}^{(j)}$ and $a^{(j)}$ are independent and the same is true
for $\widehat{c}^{(j)}$ and $a^{(j)}$ Theorem 1 implies that
\begin{equation}\label{13}
p_j^c = p_j^{\overline{c}} p_1^a + p_j^{\widehat{c}} p_2^a +
\gamma_j.
\end{equation}
So, the equality $\gamma_j = 0$ (if $\gamma_1 = 0$
then $\gamma_2 =0$ and vice versa) is satisfied if and only if the point
$p_j^c$ is appropriately located on the segment with end points
$p_j^{\overline{c}}$ and $p_j^{\widehat{c}}$.
Evidently, for any given $p_j^a$, $p_j^c$ it is always possible to indicate
$p_j^{\overline{c}}$ and $p_j^{\widehat{c}}$ $(j=1,2)$ to guarantee the
relation $\gamma_j =0$. Namely, one takes $p_j^{\overline{c}}$ in such a
way that
$$
\max\{0,(p_j^c-p_2^a)/p_1^a\}\leq p_j^{\overline{c}}\leq\min\{1,p_j^c/p_1^a\}
$$
and after that chooses
$p_j^{\widehat{c}} = (p_j^c - p_1^a p_j^{\overline{c}})/p_2^a$, $j=1,2$.

{\bf Remark 4.} It is  possible to write formula \eqref{11} analogously to
\eqref{2} using the auxiliary parameters $\lambda_j$ determined by relation
$$
\gamma_j = 2 \sqrt{p_{j1}^{\overline{c}|a} p_1^a p_{j2}^{\widehat{c}|a}
 p_2^a}\lambda_j,\;\;j=1,2.
$$
Note that in general $\lambda_j$ can take any real values. In the particular
case described by formula \eqref{13} we establish that
$$
\lambda_j \leq 1 \Longleftrightarrow  t_{j1} + t_{j2} \geq
\sqrt{p_j^c},
$$
$$
\lambda_j \geq -1 \Longleftrightarrow |t_{j1}-t_{j2}| \leq
\sqrt{p_j^c}
$$
where $t_{j1} = \sqrt{p_j^{\overline{c}}p_1^a}$, $t_{j2} =
\sqrt{p_j^{\widehat{c}}p_2^a}$ $(j=1,2)$.
If, moreover, $p_j^{\overline{c}} = p_j^{\widehat{c}}$ then
\begin{equation}\label{15}
|\lambda_j| \leq 1 \Longleftrightarrow
\frac{p_j^c}{1+2\sqrt{p_1^a p_2^a}}
\leq p_j^{\overline{c}} \leq
\frac{p_j^c}{1-2\sqrt{p_1^a p_2^a}}.
\end{equation}
The last relation shows that the possibility to represent $\lambda_j$ as
the cosine of some angle in this special case means that the distribution
of $\overline{c}^{(1)}$ is obtained from that of $c^{(1)}$ by a
"bounded perturbation".

{\bf Remark 5.} It is not difficult to complicate the scheme of measuring
characteristics of properties {\bf A} and {\bf C}.
For instance one can assume that procedures ${\cal M}_{\bf A}$ and ${\cal
M}_{\bf C}$ do not permit to fix exactly in all experiments (i.e. for all
systems) the values of characteristics for properties {\bf A} and {\bf C}.
Then for validity of Theorem 1 it is sufficient to suppose that the number of
faults (or "nonsuccessful") measurements is $P$-a.e. $o(N)$ (as $N \to
\infty$) among the systems with labels $1,\ldots,N$.  One can obtain the
analogue of Theorem 1 assuming (see e.g. Khrennikov, 1999, p.66) that prior
to measurement procedures ${\cal M}_{\bf A}$ and ${\cal M}_{\bf C}$
one has to implement some prepearing procedures ${\cal P}_{\bf A}$ and
${\cal P}_{\bf C}$ (modifying in an appropriate manner the properties of
systems
under consideration).

\vskip1cm
{\bf Aknowledgements}
\vskip0,5cm
The authors thank A.N.Shiryaev for the discussion of contextualism principle
and for the reference to the paper by Kolmogorov, 1965.
A. Khrennikov would like to thank L. Accardi, L. Ballentine, A. Holevo for
discussions on probabilistic foundations of quantum theory.
A.V.Bulinski is
grateful for hospitality to V\"axj\"o University where in May
2002 this joint work was written.

\vskip0,3cm
The research is partially supported by RFBR grant 00-01-00131 and
grant of the Swedish Royal Academy of Sciences on collablration with States
of former Soviet Union.

\vskip1cm
\begin{center}
{\bf References}
\end{center}
\vskip0,4cm

Accardi, L., 1984, The probabilistic roots of the quantum mechanical
paradoxes.
The wave--particle dualism. A tribute to Louis de Broglie on his 90th
 Birthda, (Perugia, 1982). Edited by S. Diner, D. Fargue, G. Lochak and F.
Selleri.
 D. Reidel Publ. Company, Dordrecht, 297--330.

Accardi, L. and Regoli, M., 2001. Locality and Bell's inequality.
Quantum Prob. and White Noise Analysis, 13, 1-28, WSP, Singapore.

Ballentine, L. E.,  1986. Probability theory in quantum mechanics. American
J. of Physics, 54, 883-888.

Beltrametti, E., and Cassinelli, G., 1981. The logic of quantum mechanics.
Addison-Wesley, Reading, Mass.

Bohr, N., 1934. The atomic theory and the fundamental principles underlying
the description of nature,
in The atomic theory and  the description of nature. Cambridge Univ. Press.

Borkar V.S., 1995.  Probability Theory. An advanced course. Springer, Berlin.

Busch, P.,  Grabowski,  M., Lahti,  P., 1995. Operational Quantum Physics.
Springer Verlag.

d'Espagnat,  B., 1995. Veiled Reality. An anlysis of present-day
quantum mechanical concepts. Addison-Wesley.

d'Espagnat B., 1999. Conceptual Foundations of Quantum Mechanics.
Second Edition, Perseus Books Publishing, L.L.C.

Dirac, P. A. M., 1930.  The Principles of Quantum Mechanics.
Oxford Univ. Press.

Feynman, R.  and Hibbs, A., 1965. Quantum Mechanics and Path Integrals.
McGraw-Hill, New-York.

Gnedenko, B.V., 1962.  The theory of probability. Chelsea Publ. Com.,
New-York.

Heisenberg, W.,  1930.  Physical principles of quantum theory.
Chicago Univ. Press.

Holevo, A.S., 2001. Statistical structure of quantum theory.
Lect. Notes Phys., 67, Berlin.

Holevo, A.S., 1980. Probabilistic and statistical aspects of quantum theory.
Nauka, Moscow (North Holland, 1982).

Jauch, J. M., 1968. Foundations of Quantum Mechanics. Addison-Wesley,
Reading, Mass.

Khrennikov A. Yu., 1999. Interpretations of Probability, VSP, Utrecht.

Khrennikov A. Yu., 2001a. Linear representations of probabilistic
transformations
induced by context transitions. J.Phys. A: Math. Gen. {\bf 34}, 9965--9981.

Khrennikov, A. Yu., 2001b.  Origin of quantum probabilities.
Quantum Prob. and White Noise Analysis, 13, 180-200, WSP, Singapore.

Khrennikov, A. Yu., 2000. A perturbation of CHSH inequality induced by
fluctuations
of ensemble distributions. J. of Math. Physics, 41, 5934-5944.

Kolmogoroff, A. N., 1933. Grundbegriffe der Wahrscheinlichkeitsrechnung.
Springer Verlag, Berlin; reprinted: 1956.
Foundations of the Probability Theory.
Chelsea Publ. Comp., New York.

Kolmogorov A.N., 1965. The Theory of Probability. In: A.D.Alexandrov,
A.N.Kol\-mo\-go\-rov, M.A.Lavrent'ev (Eds.) Mathematics, Its Content, Methods,
and Meaning, v.2, M.I.T. Press.

Meyer, P-A., 1993. Quantum probability for probabilists. Lecture Notes in
Math.,
1538, Heidelberg.

Parthasaraty, K. R., 1992. An introduction to quantum stochastic calculus.
Birkh\"auser, Basel.

Peres, A., 1994. Quantum Theory: Concepts and Methods. Kluwer Academic
Publishers.

Shiryaev A.N., 1998.  Mathematical Probability Theory. Essays of Formation
History.  Supplement to A.N.Kolmogorov "Foundations of Probability
Theory" (3 ed.). Fazis, Moscow (in Russian).

Thorisson H., 2000. Coupling, Stationarity and Regeneration. Springer, Berlin.

von Mises, R. , 1964.  The mathematical theory of probability and
statistics. Academic, London.

von Neumann,  J., 1955. Mathematical foundations
of quantum mechanics. Princeton Univ. Press, Princeton, N.J.
\end{document}